# Possible Size Dependence of Distribution Functions of Classic, Boson, and Fermion Assemblies

Mikrajuddin Abdullah[(1,2,a)]

[(1)]Department of Physics Bandung Institute of Technology

Jalan Ganesa 10 Bandung 40132, Indonesia

[(2)]Bandung Innovation Center

JalanSembrani 19 Bandung, Indonesia

[(a)]E-mail: mikrajuddin@gmal.com

**Abstract**

I derived the size dependence of distribution function for classic, boson, and fermion assemblies. I did not use the Stirling approximation so that deviation contributed by this approximation at small number of particles can be avoided. I identified that the size dependence of the distribution function is contained in the fermi energy or chemical potential. My results seem to match few reports on the dependence of fermi energy or chemical potential on particle size of several nanometer sized materials.

## I. INTRODUCTION

The Maxwell-Boltzmann, Bose-Einstein, and Fermi-Dirac distribution functions have been commonly derived using method of permutation of particles in groups of energies in assemblies [1-3]. The permutation is used to determine the number of configuration (number of ways of different arrangements of particles). The particle arrangements that produces the maximum configuration is taken to be a representation the macroscopic state [1-3]. The number of configuration is usually expressed in factorial forms. For large number of particles composing the assembly, i.e. approximates the Avogadro number, the permutation expression might be approximated using a Stirling formula $\ln n_i! \approx n_i \ln n_i - n_i$. Having taken this approximation, the next step is to identify number of particles occupying each group that produces the maximum configuration by applying two constraints: energy and number of particles conservations. Using the standard mathematical steps, three distribution functions (Maxwell-Boltzmann, Bose-Einstein, and Fermi-Dirac) can be obtained simply.

As explained above, the three distribution functions have been derived by assuming the number of particles composing the system is very large so that the Stirling approximation produces accurate result. However, for small number of particles, the Stirling approximation deviates and the error becomes larger for very small number of particles. As illustration, **Figure 1** shows the relative error of Stirling approximation for ln *N*! to exact value of ln *N*!, defined as

$$\eta = \frac{\ln N! - (N \ln N - N)}{\ln N!} \times 100\% \qquad (1)$$

It is apparent that the deviation is appreciable for *N* less than 200 and insignificant for *N* of larger than 200. The challenging question is, if number of particle composing an assembly is in the order of hundreds, are the well-known Maxwell-Boltzmann, Bose-Einstein and Fermi-Dirac distribution functions still valid or we should have different form of distribution functions? In other words, does the distribution function depend on the assembly size (number of particles composing the assembly)? This question might be illustrated in **Figure 2**.

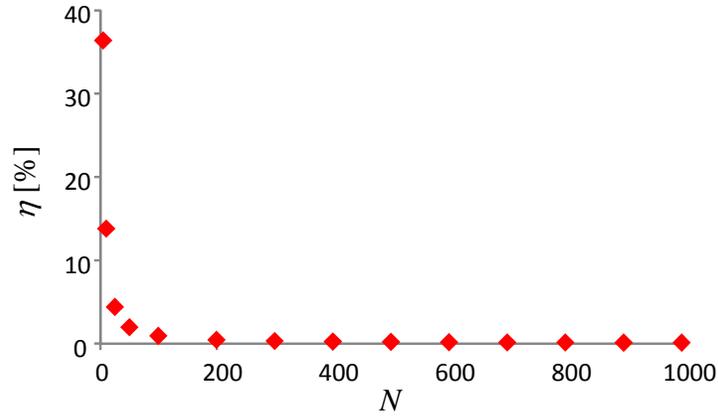

**Figure 1**. The relative error of Stirling approximation for ln $N$! to exact value of ln $N$! as defined in Eq. (1)

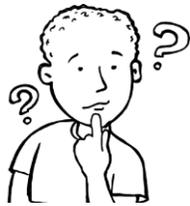
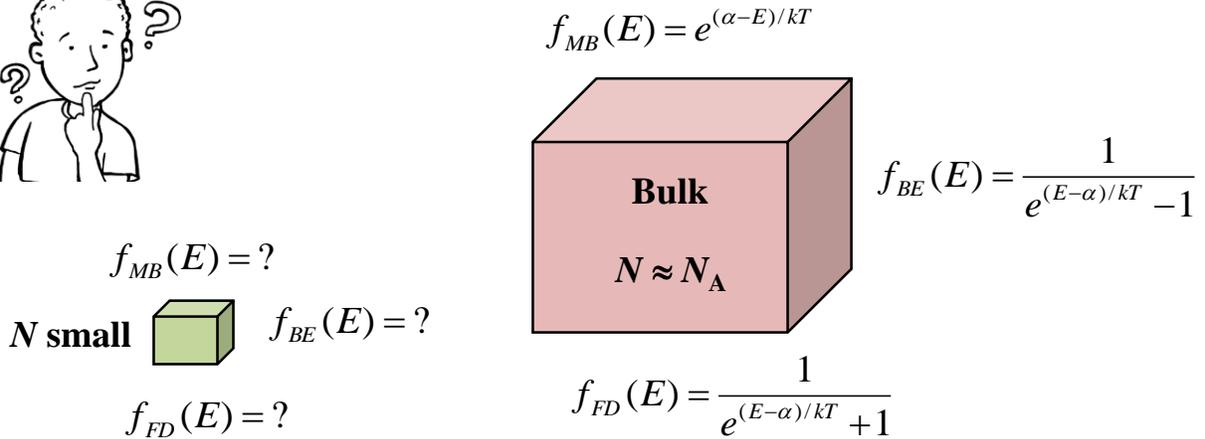

**Figure 2.** The question regarding the expression for distribution functions when the assembly size is very small. Are the expressions still the same as the well-known expressions or we must have different forms of expressions?

This question becomes interesting since in researches related to nanostructure materials, we are dealing with particles of very small number, either atoms of electrons. Recently, the behaviors of materials consisting only tens of atoms have been reported. For example,

Uppenbrink and Wales investigated the clusters of 13-150 Lennard-Jones atoms and found that different capping layers could greatly affect the stability of the particles [4]. Wales and Munro further studied the morphology change between icosahedra, cuboctahedra, and decahedra in metal clusters with 13, 55, and 147 atoms (magic number) [5]. Yang et al explored NP shape fluctuations in a single isolated Ni NP having number of atoms down to 55 [6].

At the same time, the researchers apply the distribution function when explaining several properties of the materials. For example, when describing the electrical conductivity and Seebeck coefficient for low dimensions materials, the commonly used formulae are $\sigma = \int_0^\infty \sigma(E)(-\partial f / \partial E) dE$ and $S = (k_B / q\sigma) \int_0^\infty \sigma(E)[(E - E_F)/k_B T](\partial f / \partial E) dE$ [7], should we use the distribution function that have been derived from large number of particles (bulk) or we must use a modified distribution functions?

The objective of this work is to derive the distribution functions for assembly consisting arbitrary number of particles include very small number of particles. I will derive the distribution function for classic, boson, and fermion assemblies. To avoid the use of Stirling approximation that proven to deviate for small number of particles I use the similar method as discussed by Turoff [8]. Using this approach we are free to apply the method for arbitrary assembly sizes (arbitrary number of particles).

## II. THEORY

Suppose we have an assembly consisting of $N$ particles, having volume $V$ at temperature $T$. We suppose the energies of the assembly are divided into groups, where the $i$-th group has average energy $\varepsilon_i$, number of states $g_i$ and occupied by $n_i$ particles as illustrated in **Figure 3**. As explained in standard statistical physics books [1-3], the number of different ways of arranging the particles into all groups are

$$W_c = N! \prod_i \frac{g_i^{n_i}}{n_i!} \qquad (2)$$

$$W_b = \prod_i \frac{(n_i + g_i - 1)!}{n_i!(g_i - 1)!} \tag{3}$$

$$W_f = \prod_i \frac{g_i!}{n_i!(g_i - n_i)!} \tag{4}$$

for classic (c), boson (b), and fermion (f) assemblies, respectively.

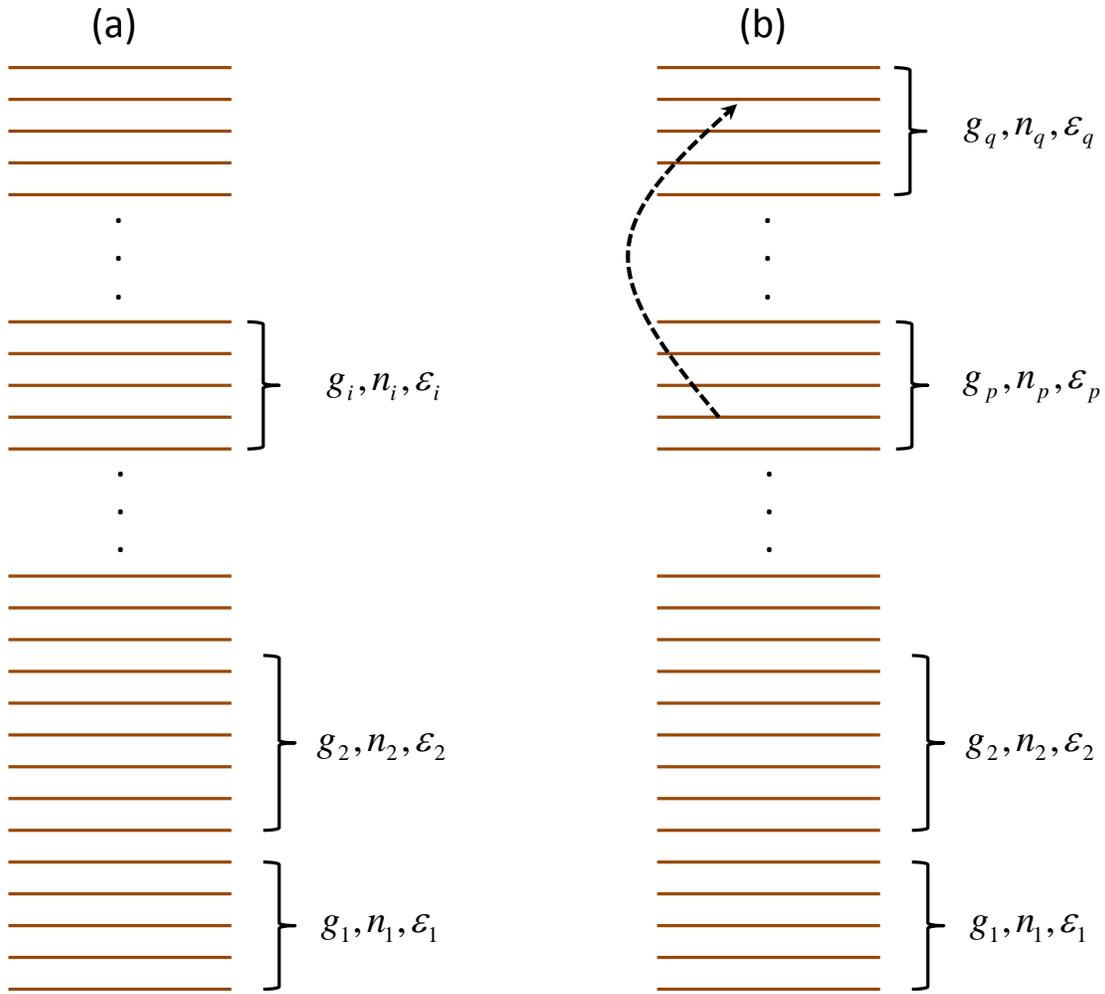

**Figure 3.** (a) The energies of particles composing an assembly are assumed to be discrete, and several nearest energies are grouped. The $i$-th group has $g_i$ states, an average energy of $\varepsilon_i$ and occupied by $n_i$ particles. (b) One particle in a state in the $p$-th group jumps to a state in the $q$-th group, implying the number of particle in the $p$-th group decreases by one while the number of particle in the $q$-th group increases by one.

Suppose an incremental energy $dU$ is given to the assembly so that one particles from a state in the $p$-th group jumps to a state in the $q$-th group (Figure 1(**b**)). The energy in the $p$-th group is $\varepsilon_p$ and in the $q$-th group is $\varepsilon_q$. Therefore, the energy incremental satisfies $dU = \varepsilon_q - \varepsilon_p$. This jumping causes the number of particles in the $p$-th ($q$-th) group changes to become $n_p - 1$ ( $n_q + 1$) so that by referring to Eqs (5)-(7), the number of different ways of arranging particles in new configurations become [8]

$$W'_c = \left( N! \prod_i \frac{g_i^{n_i}}{n_i!} \right) \left( \frac{n_p}{g_p} \frac{g_q}{n_q + 1} \right)$$

$$= W_c \left( \frac{n_p}{g_p} \frac{g_q}{n_q + 1} \right) \tag{5}$$

$$W'_b = W_b \left( \frac{n_p}{n_p + g_p - 1} \frac{n_q + g_q}{n_q + 1} \right) \tag{6}$$

$$W'_f = W_f \left( \frac{n_p}{g_p - n_p + 1} \frac{g_q - n_q}{n_q + 1} \right) \tag{7}$$

We assumed that, even the number of particles composing the assembly is finite, however the following condition is always satisfied, $n_p, n_q \gg 1$. Based on this assumed condition, Eqs (5)-(7) can be approximated as

$$W'_c \approx W_c \left( \frac{n_p}{g_p} \frac{g_q}{n_q} \right) \tag{8}$$

$$W'_b \approx W_b\left(\frac{n_p}{n_p+g_p}\frac{n_q+g_q}{n_q}\right) \tag{9}$$

$$W'_f \approx W_f\left(\frac{n_p}{g_p-n_p}\frac{g_q-n_q}{n_q}\right) \tag{10}$$

We then use the definition of entropy as $S = k\ln W_j$, $S' = k\ln W'_j$ where the index $j$ refers either to classic, boson, or fermion. The change of entropy due to a particle jumping is $dS = S' - S$. Based on this definition, Eqs (8)-(10) change to become

$$dS_c \approx k\ln\left(\frac{n_p}{g_p}\frac{g_q}{n_q}\right) \tag{11}$$

$$dS_b \approx k\ln\left(\frac{n_p}{n_p+g_p}\frac{n_q+g_q}{n_q}\right) \tag{12}$$

$$dS_f \approx k\ln\left(\frac{n_p}{g_p-n_p}\frac{g_q-n_q}{n_q}\right) \tag{13}$$

We have a thermodynamic equation relating the energy, entropy, and temperature as $T = \partial U/\partial S$ or $dU \approx TdS$. If this definition is applied to Eqs. (11) – (13) we obtain the equation for classical, boson, and fermion as

$$\varepsilon_q - \varepsilon_p \approx kT\ln\left(\frac{n_p}{g_p}\frac{g_q}{n_q}\right)_c \tag{14}$$

$$\varepsilon_q - \varepsilon_p \approx kT\ln\left(\frac{n_p}{n_p+g_p}\frac{n_q+g_q}{n_q}\right)_b \tag{15}$$

$$\varepsilon_q - \varepsilon_p \approx kT\ln\left(\frac{n_p}{g_p-n_p}\frac{g_q-n_q}{n_q}\right)_f \tag{16}$$

Whish can be rearranged as

$$\frac{n_p / g_p}{n_q / g_q} = \frac{\exp(-\varepsilon_p / kT)}{\exp(-\varepsilon_q / kT)} \tag{17}$$

$$\frac{n_p /(n_p + g_p)}{n_q /(n_q + g_q)} = \frac{\exp(-\varepsilon_p / kT)}{\exp(-\varepsilon_q / kT)} \tag{18}$$

$$\frac{n_p /(g_p - n_p)}{n_q /(g_q - n_q)} = \frac{\exp(-\varepsilon_p / kT)}{\exp(-\varepsilon_q / kT)} \tag{19}$$

From Eqs (17)-(19) we derive the following equations

$$\frac{n_p}{g_p} = \psi_c(V,T) e^{-\varepsilon_p / kT} \tag{20}$$

$$\frac{n_p}{n_p + g_p} = \psi_b(V,T) e^{-\varepsilon_p / kT} \tag{21}$$

$$\frac{n_q}{g_q - n_q} = \psi_f(V,T) e^{-\varepsilon_q / kT} \tag{22}$$

For assembly of classic, fermion, and boson particles, respectively, where $\psi_c$, $\psi_b$, and $\psi_f$ depend only on constant parameters belong to assembly. In equilibrium, the constant parameters are temperature and volume.

Let us define $\psi_j(V,T) = e^{\alpha_j(V,T)/kT}$, where $j$ can refer to classic, boson, or fermion. Based on this definition, the dependence of number of particles in the $p$-th group on the number of states in the same group can be expressed as

$$n_p^c = g_p e^{(\alpha_c(V,T) - \varepsilon_p)/kT} \tag{23}$$

$$n_p^b = \frac{g_p}{e^{(\varepsilon_p - \alpha_b(V,T))/kT} - 1} \tag{24}$$

$$n_p^f = \frac{g_p}{e^{(\varepsilon_p - \alpha_f(V,T))/kT} + 1} \tag{25}$$

Equations (23)-(25) describe the dependence of distribution function on assembly size merely because of the dependence of chemical potential on size. The form the the distribution function itself does not change. The form of the distribution function is universal for all assembly sizes. Different assembly sizes only have different chemical potential. From this argument, we readily obtain the distribution function of classical particles, boson, and fermion as

$$f_{MB}(E) = e^{[\alpha_c(V,T)-E]/kT} \tag{26}$$

$$f_{BE}(E) = \frac{1}{e^{[E-\alpha_b(V,T)]/kT} - 1} \tag{27}$$

$$f_{FD}(E) = \frac{1}{e^{[E-\alpha_f(V,T)]/kT} + 1} \tag{28}$$

## III. DISCUSSION

The dependence of chemical potential of boson on the number of particles has been well known, particularly when explaining the phenomenon of Bose-Einstein condensation. By referring to Figure 1, suppose the lowest group has only one state so that $g_0 = 1$. Therefore, the number of boson in this group (containing only one state) having energy $\varepsilon_0$ is

$$n_0^b = \frac{1}{e^{(\varepsilon_0 - \alpha_b(V,T))/kT} - 1} \tag{29}$$

The phenomenon of Bose-Einstein condensation occurs when nearly all bosons occupy the lowest group. If the number of boson is $N$ then when the Bose-Einstein condensation occurs we have $n_0^b \approx N$ so that Eq (29) becomes $N \approx [e^{(\varepsilon_0 - \alpha_b(V,T))/kT} - 1]^{-1}$ or $e^{(\varepsilon_0 - \alpha_b(V,T))/kT} = 1 + 1/N$, which implies

$$\alpha_b(V,T) \approx \varepsilon_0 - kT\ln(1+1/N) \tag{30}$$

For particles confining in an assembly with size $L$ (can be particle or rod radius or film thickness), the energy quantisize as $\varepsilon_s \propto 1/L^2$. The number of particles also depends on the assembly size according to $N \propto L^D$ with $D$ =3, 2, and 1 for particle, rod, or thin film, respectively. Therefore we can write Eq. (30) as

$$\alpha_b(V,T) \approx \frac{\kappa}{L^2} - kT\ln\left(1+\frac{\gamma}{L^D}\right) \qquad (31)$$

with κ and γ are constants that depends on material properties. For example γ depends on the cell volume or lattice parameter and κ depend of the effective mass of particles. It is clear from Eq (31) that the chemical potential depends on the number of particle in an assembly. For very large $L$ (very large assembly size) we approximate $\alpha_b(V,T) \approx \kappa/L^2 - \gamma kT/L^D$, implying $\alpha_b(V,T) \to 0$ if $L \to \infty$ (bulk). But, for very small $L$, a correction must be subjected to this zero value.

The challenging query is whether the chemical potential or fermi energy depends on the size of assembly or number of particles composing the system. Indeed, there is less reports on this behavior. Kiyonaga et al reported the increase in fermi energy in $TiO_2$-supported Au particles of the size range between 3.0 – 13 nm, where the fermi energy increases with diameter [9]. In addition, there are several theoretical works as well as simulation regarding the dependence of fermi energy or chemical potential on size [10-13]. Sieperman derived equations explaining the dependence of excess chemical potential on the number of particles composing the assembly as [13]

$$\Delta\mu_{ex}(N) = \frac{1}{2N}\frac{\partial P}{\partial \rho}\left[1 - k_BT\frac{\partial \rho}{\partial P} - \rho k_BT\frac{\partial^2 \rho}{\partial P^2}\right] \qquad (32)$$

with $P$ is the pressure and $\rho$ is the density. For nanoparticle, nanorod, and thin film we have, respectively, $N \propto R^3$, $N \propto R^2$, and $N \propto t$ where $R$ is nanoparticle of nanorod radius and $t$ is film thickness. Therefore, the excess of chemical potential for nanoparticle, nanorod, and thin film are $\Delta\mu_{ex} = A/R^3$, $\Delta\mu_{ex} = B/R^2$, an $\Delta\mu_{ex} = C/t$, respectively, with $A$, $B$, and $C$ are constants.

Grigor'eva showed the dependence of fermy energy on size or the excess of chemical potential as

$$\Delta\mu_{ex} = \mu_1 + \Delta\mu_e \qquad (33)$$

with $\mu_1 = A/L$, $\Delta\mu_e = (e^2 k_F/3)(\pi/2)^3[1+0.3/k_F L]$, and $L$ is the particle dimension [12]. Korotun showed theoretically that the Fermi energy of Al and Au nanoarticles increases with decreasing particle diameter, although the increases is not continuous, and in average, the dependence satisfies the empirical function $\varepsilon_F(d) = \varepsilon_F(\infty)(1+\gamma/d^\delta)$, with γ and δ are positive constants, d is particle diameter, and $\varepsilon_F(\infty)$ is the Fermi energy in bulk state [14].

Chemical potential in an intrinsic semiconductor can be written as [15]

$$\alpha_f = \frac{1}{2}E_g + \frac{3}{4}kT\ln\left(\frac{m_h^*}{m_e^*}\right) \tag{34}$$

with $m_h^*(m_e^*)$ is the effective mass of hole (electron). However, the band gap or semiconductor has been known to depend on size such as proposed by Brus [16]. $E_g(L) = E_g(\infty) + a/L^2 - b/L$, Liao et al have better fit the band gap opening ofGaP nanowire with $\Delta E_g = (q_1 L^2 + q_2 L + q_3)^{-1} + q_4$, where $q_i$'s are fitting parameters [17]. In general, the band gap of most of semiconductors satisfies an approximated formula

$$E_g(L) \approx E_g(\infty) + \frac{c}{L^\delta} \tag{35}$$

with δ varies between 1 to 1.4 [18-20]. Example of reported δ are 1.16 [21], 1.3 [22], and 1.37 [23]. However, large power has been observed such as for undoped ZnO and aluminium-doped ZnO films, the band gap opening as function of grain size have been well fit using δ = 1.65 [24]. From this finding, the general expression for chemical potential inside intrinsic semiconductor reads

$$\alpha_f(L) \approx \alpha_f(\infty) + \frac{\sigma}{L^\delta} \tag{36}$$

with σ = c//2, which is similar to the result of Grigor'eva [12] and Korotun [14].

For classical assemblies, the factor $e^{\alpha_c(V,T)/kT}$ nearly does not have effect due to normalization of distribution function. The total number of particles satisfies $N = \sum_p n_p^c$ or $N = e^{\alpha_c(V,T)/kT} \sum_p g_p e^{-\varepsilon_p/kT}$, resulting

$$e^{\alpha_c(V,T)/kT} = \frac{N}{\sum_p g_p e^{-\varepsilon_p/kT}} \tag{37}$$

Therefore, the number of classical particles in the p-th group can be weitten as

$$n_p^c = N \frac{g_p e^{-\varepsilon_p/kT}}{\sum_p g_p e^{-\varepsilon_p/kT}} \tag{38}$$

It is clear from Eq. (38) that the factor $e^{\alpha_c(V,T)/kT}$ is absent. We than conclude that the distribution function for classical assembly unchanges for all assembly sizes.

## IV. CONCLUSION

I have derived the size dependence of distribution function for classic, boson, and fermion assemblies. I identified that the size dependence of the distribution function is contained in the fermi energy or chemical potential. The proposed results seem to match few reports on the dependence of fermi energy or chemical potential on particle size of several nanometer sized materials. The results might be applied to explain several phenomena in nanomaterials especially the phenomena that have been derived using distribution function.

**References**

[1] A. J. Pointon, An Introduction to statistical physics for students, New York: Longman (1967).